\documentclass[final,authoryear,5p,times,twocolumn]{elsarticle}
\usepackage{amssymb}
\usepackage{gensymb}

\usepackage[pdftex,bookmarks,bookmarksopen,colorlinks]{hyperref}
\usepackage{color,soul}

\usepackage{upquote}

\usepackage{enumitem}
\usepackage{etoolbox}

\usepackage{lineno}
\usepackage{graphics}
\usepackage{graphicx}
\usepackage{amsmath}
\usepackage{algorithmicx}
\usepackage{algorithm}
\usepackage{algpseudocode}
\usepackage{afterpage}
\usepackage{url}
\usepackage{tablefootnote}
\usepackage{epstopdf}
\usepackage{mathtools}

\journal{Astronomy and Computing}

\begin{document}

\begin{frontmatter}

\title{Astronomical Imagery: Considerations For a Contemporary Approach with JPEG2000.}
 
\author[icrar]{Vyacheslav V. Kitaeff\corref{cor1}}
\ead{slava.kitaeff@icrar.org}
\author[icrar]{Andrew Cannon}
\author[icrar]{Andreas Wicenec}
\ead{andreas.wicenec@icrar.org}
\author[unsw]{David Taubman}
\ead{d.taubman@unsw.edu.au}

\cortext[cor1]{Corresponding authors}

\address[icrar]{International Centre for Radio Astronomy Research, University of Western Australia, M468, 35 Stirling Hwy, Crawley 6009, WA, Australia}
\address[unsw]{School of Electrical Engineering and Telecommunications, The University of New South Wales, UNSW, Sydney, NSW 2052, Australia}

\begin{abstract}
The new wide-field radio telescopes, such as: ASKAP, MWA, LOFAR, eVLA and SKA; 
will produce spectral-imaging data-cubes (SIDC) of unprecedented size -- in the order 
of hundreds of Petabytes. Servicing such data as images to the end-user in a traditional 
manner and formats is likely going to encounter significant performance fallbacks. We discuss the 
requirements for extremely large SIDCs, and in this light we analyse the applicability of 
the approach taken in the JPEG2000 (ISO/IEC 15444) standards. We argue the case for the adaptation 
of contemporary industry standards and technologies vs the modification of legacy astronomy 
standards or the development new from scratch.
\end{abstract}

\begin{keyword}
data format \sep
File formats \sep
JPEG2000 \sep
Standards
\MSC 85-04
\end{keyword}

\end{frontmatter}
  
\section{Introduction}

Spectral-imaging data-cubes (SIDCs), from the new radio telescopes that are currently in various stages of construction or 
commissioning -- Australian Square Kilometre Array Pathfinder (ASKAP) \citep{ASKAP..09}, 
Murchison Widefield Array (MWA) \citep{TINGAY}, LOFAR \citep{2013A&A...556A...2V}, 
MeerKAT~\citep{2009arXiv0910.2935B}, eVLA ~\citep{2011ApJ...739L...1P}
-- are expected to be in the range of tens of GBs to several TBs. The Square Kilometre 
Array (SKA) Design Reference Mission, SKA Phase 1 \citep{SPDO2011}, defines at least 
one survey, namely the ``Galaxy Evolution in the Nearby Universe: HI Observations", for 
which the SKA pipeline will produce hundreds of SIDCs, of tens of terabytes each. In its first 
year the SKA Phase 1 is expected to collect over 8 EB of data. The data volumes for the full 
SKA are expected to be by at least an order of magnitude larger.

Even taking into account projected advances in HDD/SSD and network technologies, such 
large SIDCs cannot be processed or stored on local user computers. Most of the imaging 
data will be never seen by a human, but rather processed automatically \citep{2012MNRAS.421.3242W,
2012PASA...29..318P, 2012PASA...29..352J, 2012PASA...29..371W}. However, 
there will still be a number of cases where visualisation is going to be 
required, e.g. data quality control/assessment or detailed studies of individual objects. 

Visual exploration of such large data volumes requires a new paradigm for the generation 
and servicing of the higher level data products to the end-user. In this paper 
we present a straw man of the functionality required to enable working with extremely 
large radio astronomy imagery. We consider the JPEG2000 industry 
standard as a suitable example that addresses many similar requirements, even though it was originally 
developed for medical and remote sensing imagery.

Currently, most radio astronomy imaging data is stored and distributed in one of three 
formats: FITS (Flexible Image Transport System) \citep{2010A&A...524A..42P}; 
CASA Image Tables~\footnote{\href{http://ascl.net/1107.013}{http://ascl.net/1107.013}} and newly developed by 
LOFAR HDF5-based format \citep{2011ASPC..442...53A}. 
FITS and HDF5 are, in general, single self-describing files containing the image data, as 
well as metadata. CASA, on the other hand, uses a different approach representing any 
data as a hierarchical structure of directories and files. CASA data is usually distributed 
as an archived file created by using common archiving software, such as 
\texttt{tar}\footnote{\href{http://en.wikipedia.org/wiki/Tar(computing)/}{http://en.wikipedia.org/wiki/Tar(computing)/}}. These formats provide 
both, portability and access to image data. Currently, 
image files or CASA tar-balls are normally retrieved from an archive and stored on a local 
computer for exploration, analysis or processing purposes. Alternatively, a specified part 
of an image-cube (cutout) is produced in one of the image formats, and presented to the 
user as a download. If coterminous regions are required, several cutout files would be 
produced and downloaded. The example of such a framework is Simple Image Access Protocol (SIAP)\footnote{\href{http://www. ivoa.net/Documents/SIA/}{http://www. ivoa.net/Documents/SIA/}}
 of the International Virtual Observatory Alliance (IVOA)\footnote{\href{http://www.ivoa.net}{http://www.ivoa.net}} that provides a 
uniform interface for retrieving image 
data from a variety of astronomical image repositories. 
By using SIAP the user can query compliant archives in a standardised manner and 
retrieve image files in one or more formats, depending on the archive capabilities (e.g. FITS,
PNG or JPEG). The resulting files can then be stored on a local computer or a virtual network 
storage device that is provided through VOSpace, which is another IVOA standard.

In the paper we discuss the use case of extremely large SIDCs in the context
of the limitations of the current standard astronomy file formats. We present the analysis 
of the applicability of the approach taken in developing JPEG2000 standards to addressing 
the new requirements of extremely large astronomical imagery. We also present some interesting
benchmarks from using JPEG2000 on large radio astronomy images.

The rest of paper is structured as follows. In Section~\ref{cha:use-case}, we discuss the specific requirements
of extremely large imaging. Section~\ref{cha:JPEG2000} discusses JPEG2000 standards, and
how they have addressed the requirements of extremely large imaging. We specifically discuss the image interaction
protocol in detail as the alternative to the used in astronomy cutout framework. Section~\ref{cha:Bench1}
presents benchmarks for JPEG2000 compression for radio astronomy images. In Section~\ref{cha:adopting}
we discuss the strategic approaches for improving the existing astronomy standards or
the adoption of new industry standards. Finally, we conclude in Section~\ref{cha:conc}.

\section{Use case for extremely large images} \label{cha:use-case}

\textit{ASKAP Science Data Archive: Requirements and Use Cases}\footnote{\href{http://www.atnf.csiro.au/management/atuc/2013dec/docs/ASKAP\_SW\_0017\_v0.8\_fordistribution.pdf}{CSIRO: ASKAP Science Data Archive: Requirements and Use Cases}} 
indicates the individual data product sizes up to ~2.24TB for some science cases at the maximum interferometer 
baseline 6 km. By 2020 ASKAP will be incorporated into the SKA1-Survey increasing the number of antennas from 36 to 96 
and the maximum baseline to 50 km. The individual data products can be expected as large as 32TB.

\begin{figure}
  \centering
  \includegraphics[width=90mm]{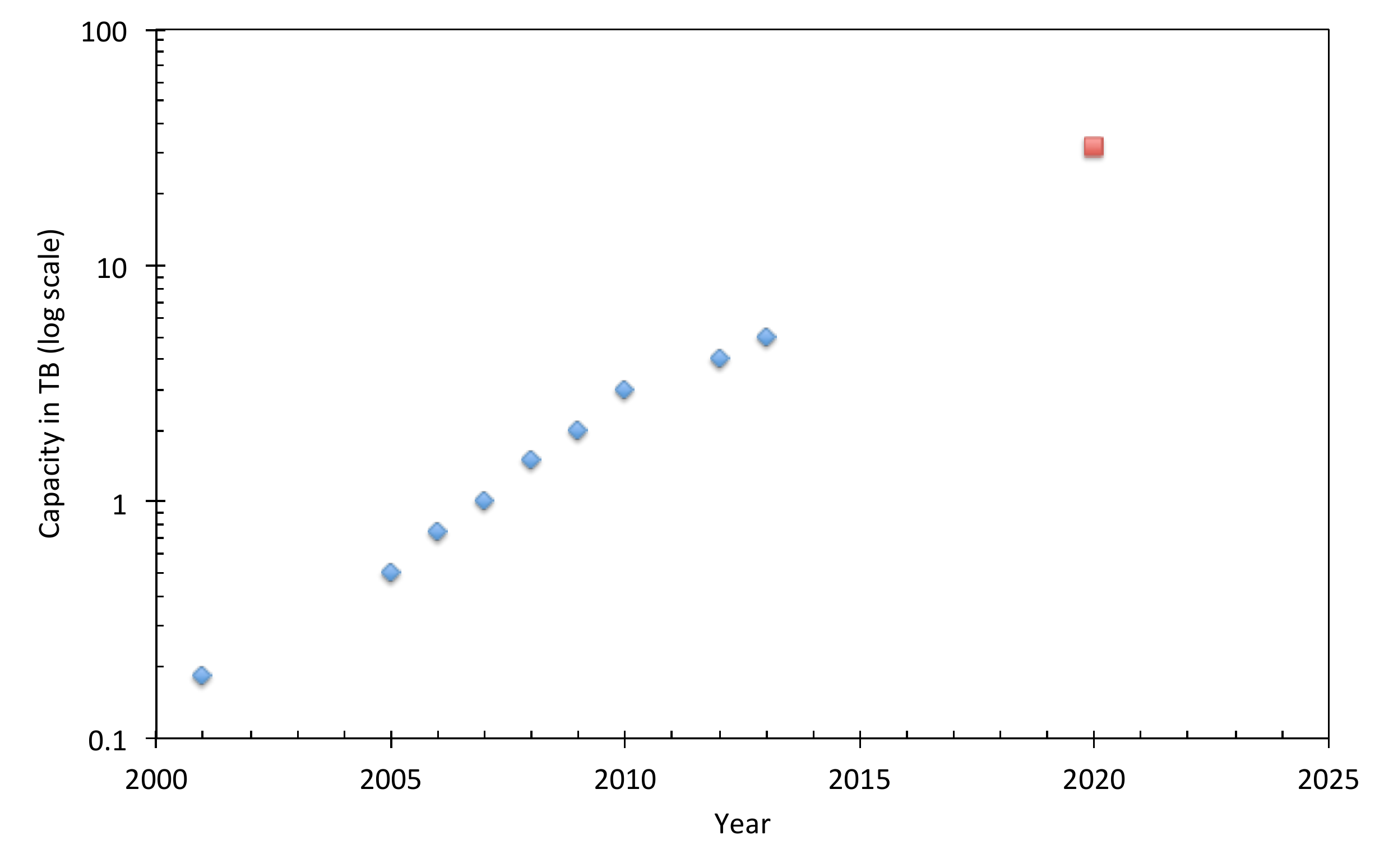}
  \caption{Historical (blue  diamonds) and projected (red squares) HDD capacity. Based on \href{http://www.storagenewsletter.com/rubriques/hard-disk-drives/milestones-in-hdd-capacity/}{http://www.storagenewsletter.com/rubriques/hard-disk-drives/milestones-in-hdd-capacity/} (accessed on 6/05/2014).}
  \label{fig:HDD-capacity}
\end{figure}

Figure~\ref{fig:HDD-capacity} shows the capacity of HDD over the years. The projection assumes 
the currently observed average growth of disc capacity at 32\% rate. The disc capacity increases
by factor of 20 every 10 years. If the same rate is sustained, by the time SKA1 is constructed 
(2020-23), the individual disc capacity can be expected to be about 32TB. 

\begin{figure}
  \centering
  \includegraphics[width=90mm]{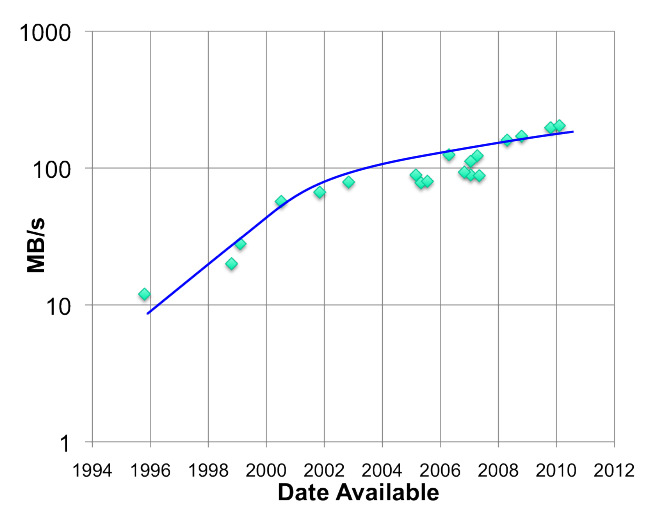}
  \caption{Maximum sustained bandwidth of HHDs over the years. Adopted from \citealt{Freitas2011}.}
  \label{fig:Max-sustained-bandwidth}
\end{figure}

\begin{figure}
  \centering
  \includegraphics[width=90mm]{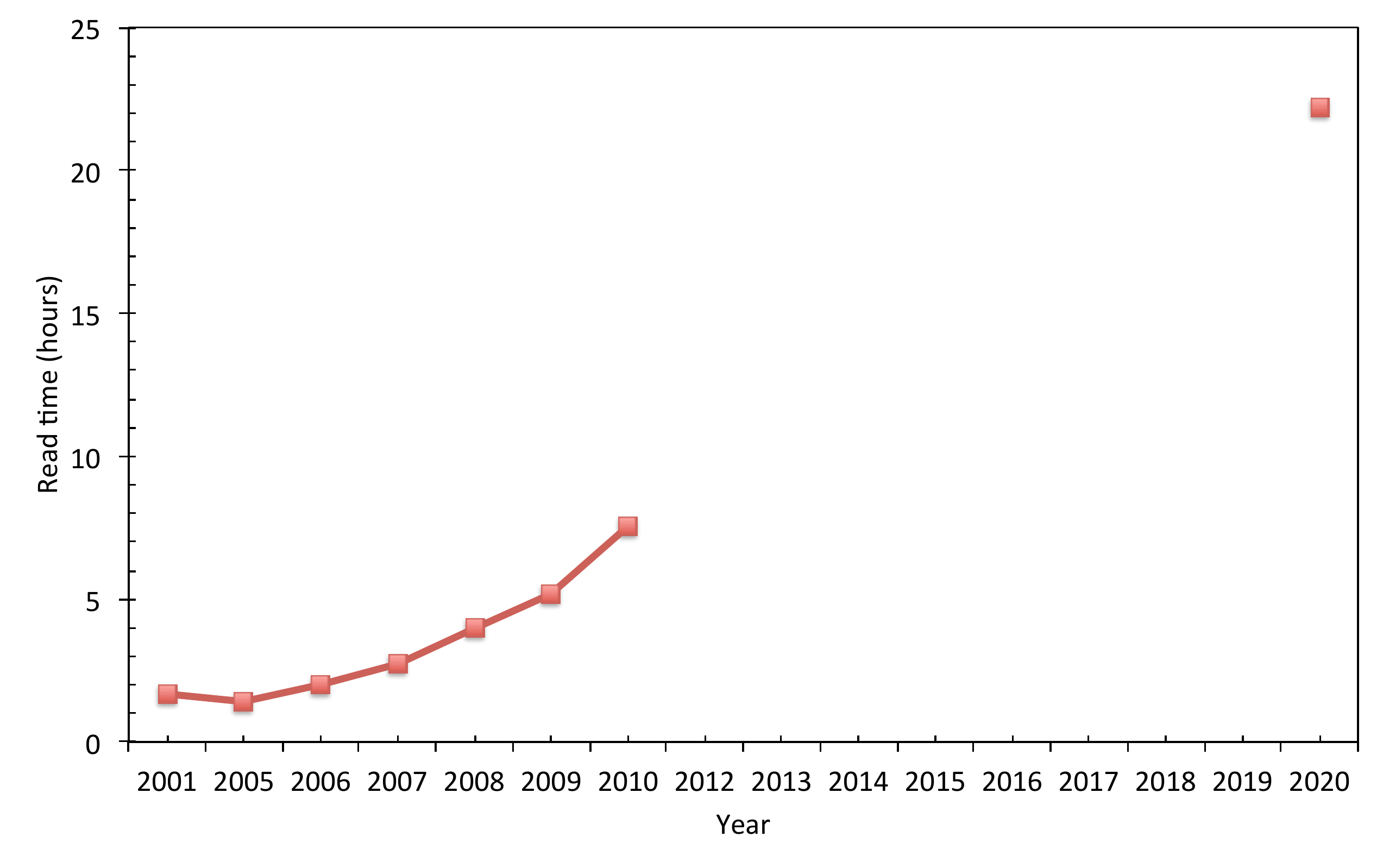}
  \caption{Read time of the entire HDD over the years with the projected time for 2020.}
  \label{fig:hdd-read-time}
\end{figure}

Figure~\ref{fig:Max-sustained-bandwidth} shows the maximum sustained bandwidth of HHDs over the years \citep{Freitas2011}.
One can see that the improvement rate is rather moderate, about 4--5 times per decade. Figure~\ref{fig:hdd-read-time} shows the increasing read time of the entire HDD over the years. 

\begin{figure}
  \centering
  \includegraphics[width=90mm]{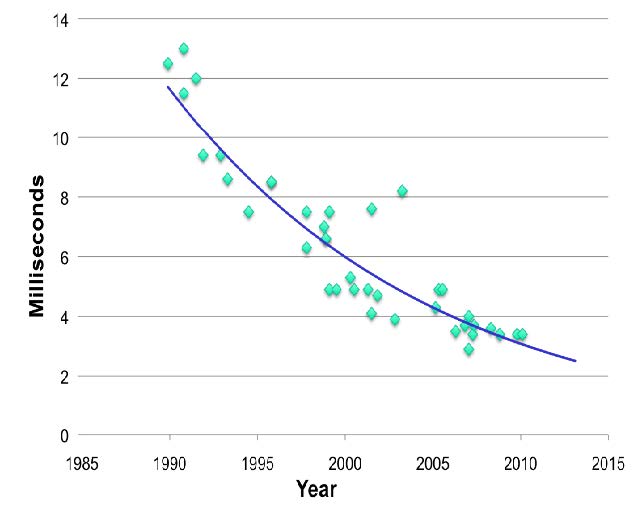}
  \caption{Average seek time trend. Adopted from \citealt{Freitas2011}.}
  \label{fig:average-seek-time}
\end{figure}

Of course, such a read only indicates the time need to read the data sequentially. In many cases,
during the scientific data analysis, the data is accessed randomly. Been a mechanical device, HDD requires a time
to relocate the head to the required position, and wait until the disc turns into the right position before the needed data 
can be accessed. This delay translates into a latency when the data need to be accessed randomly. Figure~\ref{fig:average-seek-time} shows the average seek time trend in HDD over the years. The improvement is very moderate, factor about 1.7 per decade. 

Solid State Drives (SSD) is a promising technology that may help to overcome the I/O 
bandwidth and latency problems in the future, though, at the time of writing this paper, 
the market only offers 1TB SSD for the desktop/laptop computers, while the
largest HDD is 6TB.

These all means that working with the increasing in size datasets is likely going to be 
increasingly difficult on personal computers if feasible at all. The software technologies allowing
an interactive work with the data stored on a server that can provide a fast parallel access
to the data are going to be important for the projects like SKA.

In many cases, images are not required at their full resolution or fidelity. It should be possible 
to access images at any of a multitude of reduced resolutions and/or reduced fidelities, according 
to need, all from a single master image. Such a multi-resolution representation should not lead to 
increased storage requirements, which are already high. For example, pyramid representations \citep{adelson1984pyramid}
possess the desired multi-resolution accessibility attributes, but inevitably expands the data.

We also note that not all of the image might be required at the same level of quality. Particular 
regions of interest (ROI), e.g. containing an object to be studied, such as a galaxy or a nebula, may 
need to be of much higher quality or resolution than others. Producing a cutout or many cutouts 
is a limited solution, as cutouts completely remove the surrounding area.  This is problematic 
because the surrounding area provides context for reconstructing the relationship between 
multiple objects in the field of view; moreover, the imagery within the surrounding area may 
be of interest in its own right. A much better approach, in this case, would be to have an adaptively 
encoded image, in which the regions of interest are encoded with higher fidelity/resolution than 
the surrounding areas.

Even combining such advanced techniques as multiple resolution/fidelity and adaptive 
encoding/transferring of ROI, the images can still be very large and require time to be 
transferred to the client. In the case of visual exploration of data, it would make sense to 
immediately transfer only the data that is required for display. Other parts of an image could 
be requested and transferred on demand. The protocol should be intelligent enough for such a use case.

It would also be very useful to support the progressive transfer of an image from a server to the client. That is, the user should be 
able to see the whole image of the selected region queried as soon as a first portion of the 
data is transferred, while each successive portion of the data that is transferred should serve 
to improve the quality of the displayed imagery.  By contrast, many ``pyramid" techniques 
possess only multi-resolution access, without progressive transfer, so that higher quality 
representations must completely replace the lower quality ones, leading to substantial inefficiencies 
and much higher transfer bandwidths. The client-server framework should be intelligent enough 
not to transfer more data than is necessary for displaying or processing the content that is of interest.

Further, we will demonstrate that radio astronomy imaging data can be effectively compressed, 
and the error due to the compression can be controlled. Compression significantly reduces the cost of storage, 
operations and network bandwidth. However, it should be possible to access image regions, 
resolutions and qualities directly from the compressed representation.  If the imagery must 
first be decompressed, and then re-compressed to address a users needs, this will place 
unreasonable computational and memory demands upon the server, leading to a large 
latency in service time and limited ability to serve a variety of users. Ideally, decompression 
should occur only at the point where an image is to be displayed or used. Some usage cases 
can expect large ratios in compression; examples include visual data exploration, draft mosaicing, 
etc.  Other use cases may be less tolerant to the loss of fidelity in the data, e.g. source finding. It 
follows that multiple levels of compression should be available: 
\begin{itemize}
\item high fidelity, potentially even numerically lossless compression, in which the decompressed 
image is either an exact reproduction of the original uncompressed image, or differs by 
considerably less than the intrinsic uncertainties in the imaging process; and
\item lossy compression, where the decompressed image exhibits higher levels of distortion that 
are considered acceptable in exchange for corresponding reductions in communication bandwidth 
or storage requirements.
\end{itemize}

As is suggested by the last point above, distortion metrics need to be defined and made available 
to a user, so that the impact of lossy compression can be controlled.  Such metrics involve: 
\begin{itemize}
\item statistical characterisation of how the decompressed image can be expected to differ 
from the original image; and
\item measures of the impact that different levels of distortion can be expected to have on some 
specific purposes of data exploration, e.g. source finding.
\end{itemize}
The second point is especially important, given that much of the new Radio Astronomy science is done at a very low 
signal-to-noise ratio (SNR).

\section{Case study: JPEG2000} \label{cha:JPEG2000}
In developing a contemporary protocol for working with extremely large astronomical images it is 
useful to study how other communities have approached this problem. Indeed, large images are 
not unique to astronomy, though new telescopes such as the SKA will be at the very extreme end of 
the spectrum. Medical imaging, remote sensing, geographic information systems, virtual microscopy, 
high definition video and other applications have long histories of development in the imaging domain. 
The large size of images is not the only similarity. Multi-frequency, multi-component, volumetric data 
sets, and metadata are common attributes in a range of existing imaging fields. A number of advanced 
image/metadata formats and access layer protocols have been developed over the years\footnote{\href{http://en.wikipedia.org/wiki/Comparison\_of\_graphics\_file\_formats}{http://en.wikipedia.org/wiki/Comparison\_of\_graphics\_file\_formats}}. Some of the formats use wavelet encoding
that enables not only efficient compression but also advanced options for interaction with the image
data (e.g. MrSID\footnote{\href{http://en.wikipedia.org/wiki/MrSID}{http://en.wikipedia.org/wiki/MrSID}}, JPEG2000\footnote{\href{http://www.jpeg.org/jpeg2000/}{http://www.jpeg.org/jpeg2000/}}, or ECW\footnote{\href{http://en.wikipedia.org/wiki/ECW\_(file\_format)}{http://en.wikipedia.org/wiki/ECW\_(file\_format)}}).

One of those, namely JPEG2000, has been developed into a comprehensive royalty free industry standard -- ISO/IEC 15444.
Due to the specific focus of the standard on the large imagery, instead of the consumer photography, 
the standard has become widely adopted by the industries, such as medical imaging \citep{1028146}, 
meteorology and remote sensing \citep{1294863}, Sun \citep{JHelioviewer} and planetary imaging \citep{Powell2010}, 
microscopy~\citep{2014Microscopy}, etc. We believe that the astronomy community may benefit from this development,
and learn from those industries that had faced similar challenges before astronomy.

JPEG2000 is an image compression standard and coding system created by the Joint Photographic 
Experts Group committee and published as the international standard JPEG2000 \citep{Taubman02} in 2000. 
The standard was developed to address weaknesses in existing image compression standards and 
provide new features, specifically addressing the issue of working with large images. Considerable 
effort has been made to ensure that the JPEG2000 codec can be implemented free of royalties. 
Today, there is a growing level of support for the JPEG2000 standard, through both proprietary and 
open source software libraries such as: OpenJPEG\footnote{\href{http://www.openjpeg.org}{http://www.openjpeg.org}}, JasPer\citep{JasPer}, 
Aware\footnote{\href{http://www.aware.com/imaging/jpeg2000sdk.html}{http://www.aware.com/imaging/jpeg2000sdk.html}}. JPEG2000 has been successfully used 
in a number of astronomy applications already, including the HiRISE (high resolution Mars imaging) 
project \citep{Powell2010} and JHelioviewer (high resolution Sun images) \citep{JHelioviewer}. 

The following key objectives were considered during the development of the standard. It was expected 
to allow efficient lossy and lossless compression within a single unified coding framework as well as to 
provide superior image quality, both objectively and subjectively, at 
high and low bit rates. It was expected to support additional features such as: ROI coding, a more 
flexible file format, and, at the same time, to avoid excessive computational and memory complexity, 
and excessive need for bandwidth to view an image remotely.

The main advantage offered by the approach used in JPEG2000 is the significant flexibility of its 
codestream. The codestream obtained after compression of an image with JPEG2000 is scalable, 
meaning that it can be decoded in a number of different ways.  For instance, by truncating the 
codestream at any point, a lower resolution or signal-to-noise ratio representation of the image 
can be attained; moreover, the truncated representation remains efficient, in terms of the tradeoff 
that it represents between fidelity and compressed size.  By ordering the codestream in various 
ways, applications can exploit this so-called ``scalability" attribute to achieve significant performance 
benefits \citep{Taubman02}.

The following main features of JPEG2000 make it an attractive approach for astronomy:
\begin{itemize}[leftmargin=*]
\item High compression performance, substantially superior to JPEG. 
\item Availability of multi-component transforms, including arbitrary inter-component wavelet transforms 
and arbitrary linear transforms (e.g., KLT, block-wise KLT, etc.), with both reversible and irreversible versions.
\item Multiple resolution representation.
\item Progressive transmission (or recovery) by fidelity or resolution, or both.
\item Lossless and lossy compression in a single compression architecture. Lossless compression 
is provided by the use of a reversible integer wavelet transform and progressive transmission of a lossless 
representation provides lossy to lossless refinement.
\item Random codestream access and processing, also identified as \emph{ROI: JPEG2000} codestreams, offer 
several mechanisms to support spatial random access to regions of interest, at varying degrees of granularity. 
These allow different parts of the same picture to be stored and/or retrieved at different quality levels.
\item Error resilience -- JPEG2000 is robust to bit errors introduced by communication channels, due to 
the coding of data in relatively small independent blocks within the transform domain.
\item Flexible file format -- The JPX file format, in particular, allows for rich and flexible description 
and composition of components. It allows images to be composed from any number of independently 
compressed codestreams.
\item Extensive metadata support and handling.
\item Support for volumetric image cubes, either through the specific set of extensions in Part 10 (a.k.a. ``JP3D") 
or by using the extensive set of multi-component transforms provided with Part 2 of the standard.
\item Interactivity in networked applications, as developed in the JPEG2000 Part 9 JPIP protocol.
\end{itemize}

\subsection{Encoding/decoding}

\begin{figure}
  \centering
  \includegraphics[width=90mm]{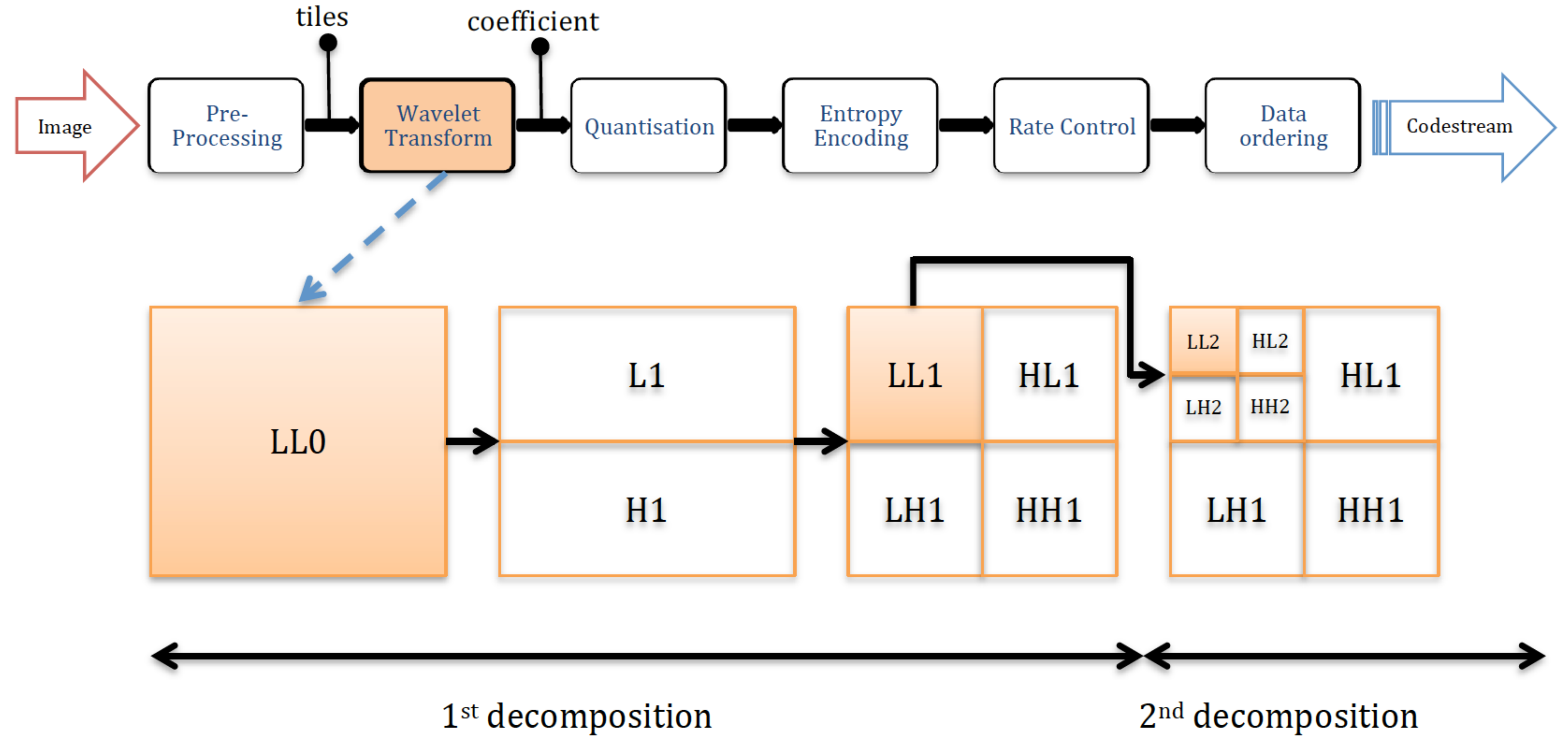}
  \caption{JPEG2000 encoding is based on discrete wavelet transformation, scalar quantisation,
context modelling, arithmetic coding and post-compression rate allocation.}
  \label{fig:jpeg2000-encoding}
\end{figure}

Unlike the binary compression available through \texttt{cfitsio} or HDF5, JPEG2000 is a true image
compression that takes advantage of the multidimensionality of data. Figure~\ref{fig:jpeg2000-encoding} depicts 
the stages of encoding in JPEG2000.

In the first stage, pre-processing is performed. Pre-processing actually contains three substages: 
Tiling, Level Offset, Reversible/Irreversible Color Transform. This stage prepares the data to correctly perform
the Wavelet Transform. During the Wavelet Transform, image components are passed recursively through the 
low pass and high pass Wavelet filters. This enables an intra-component decorrelation that
concentrates the image information in a small and very localised area. It enables the multi-resolution
image representation. The result is that 4 sub-bands with the upper left one $LL$ on Figure~\ref{fig:jpeg2000-encoding} containing
all low frequencies (low resolution image), $HL$ containing vertical high frequencies, LH containing horizontal high frequencies, and $HH$
containing diagonal high frequencies. Successive decompositions are applied on the low
frequencies $LL$ recursively as many times as desired.

By itself the Wavelet Transform does not compress the image data; it restructures the image information so
that it is easier to compress. Once the Discrete Wavelet Transform (DWT) has been
applied, the output is quantified in Quantisation unit.

Before coding is performed, the sub-bands of each tile are further partitioned into small code-blocks (e.g. 64x64
or 32$x$32 samples) such that code blocks from a sub-band have the same size. Code-blocks
are used to permit a flexible bit stream organisation.

The quantised data is then encoded in the Entropy Coding unit.
The Entropy Coding unit is composed of a Coefficient Bit Modeller and the Arithmetic Coder itself.
The Arithmetic Coder removes the redundancy in the encoding of the data. It assigns short code-words to
the more probable events and longer code-words to the less probable ones.
The Bit Modeller estimates the probability of each possible event at each point in the coding stream.

At the same time as embedded block coding is being performed, the resulting bit streams
for each code-block are organised into quality layers. A quality layer is a collection of some
consecutive bit-plane coding passes from all code-blocks in all sub-bands and all components, 
or simply stated, from each tile. Each code-block can contribute an arbitrary number
of bit-plane coding passes to a layer, but not all coding passes must be assigned to a quality
layer. Every additional layer successively increases the image quality.

Once the image has been compressed, the compressed blocks are passed over to the Rate Control unit 
that determines the extent to which each block's embedded bit stream should be truncated in order to achieve 
the target bit rate. The ideal truncation strategy is one that minimises distortion while still reaching the target bit-rate.

In Data Ordering unit, the compressed data from the bit-plane coding passes are first separated into packets. One packet is generated for each precinct in a tile. A precinct is
essentially a grouping of code blocks within a resolution level. Then, the packets are multiplexed 
together in an ordered manner to form one code-stream. There are five built-in ways to order the packets, called 
progressions, where position refers to the precinct number:
\begin{itemize}[leftmargin=*]
\item Quality: layer, resolution, component, position
\item Resolution 1: resolution, layer, component, position
\item Resolution 2: resolution, position, component, layer
\item Position: position, component, resolution, layer
\item Component: component, position, resolution, layer
\end{itemize}

The decoder basically performs the opposite operations of the encoder.

The details and mathematics of JPEG2000 encoding can be found in \citealt{Gray}, \citealt{Adams2001}, or \citealt{Li2003}.

\subsection{File format and metadata}

\begin{figure}
  \centering
  \includegraphics[width=90mm]{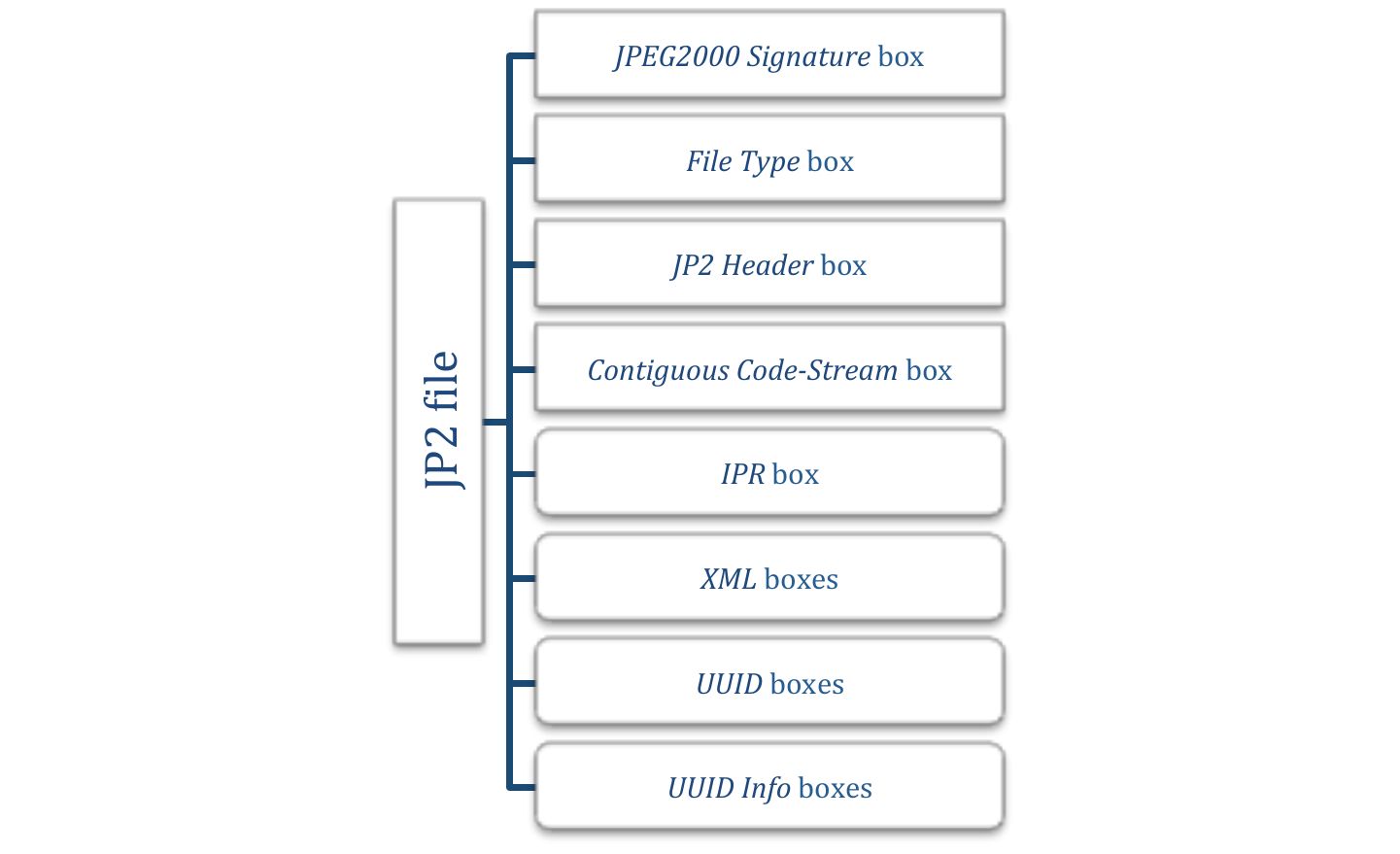}
  \caption{JP2 file format structure. Rounded conners indicate optional boxes.}
  \label{fig:jp2-file}
\end{figure}

The \textit{JP2} file format is organised as a sequence of "boxes", as depicted
in Figure~\ref{fig:jp2-file}. Boxes play a role in the file format similar to that of marker segments 
in the code-stream syntax, and they appear consecutively in the file.

There are four required required boxes: \textit{JPEG200 Signature}, \textit{File Type}, \textit{JP2 Header}, 
and \textit{Contiguous Code-Stream} boxes.

\textit{IPR}, \textit{XML}, \textit{UUID}, and \textit{UUID Info} boxes are all optional and may appear in 
any order, anywhere after the \textit{File Type} box. There may be multiple instances of these three boxes.

The \textit{JPEG2000 Signature} box identifies the file as belonging to the JPEG2000 family of file formats. 
The \textit{File Type} box identifies the file specifically as a JP2 file. The \textit{JP2 Header} box contains information 
such as image size, bit-depth, resolution, and colour space. The \textit{Contiguous Code-Stream} box contains 
a single valid JPEG2000 code-stream. \textit{IPR contains} Intellectual Property Rights information. \textit{XML} boxes 
provide for the inclusion of additional structured information, while \textit{UUID} and \textit{UUID} Info boxes provide 
a mechanism for defining vendor specific extensions.

Each of the boxes has an internal structure and sub-boxes containing the information about the image. The
details can be found in e.g. \citealt{Taubman02}.

The \textit{XML} box may contain any information whatsoever, provided that it complies to the XML 
(extensible Markup Language). For example, the discussed later \texttt{SkuareView} software uses \textit{XML} 
box to contain FITS header "as is" wrapped in a simple XML envelop. Alternatively, one of the IVOA's data models 
or some proprietary custom information could be placed in a single or multiple XML boxes.

The \textit{JPX} file format provides even more advanced metadata handling \citep{15444-2}.

\subsection{JPIP}
JPIP protocol deserves a special consideration as it offers significantly richer functionality compared to IVOA SIAP.

JPIP (JPEG2000 Interactive Protocol)) is a client/server communication protocol that 
enables a server to transmit only those portions of a JPEG2000 image that are applicable to 
the client's immediate needs. However, this is achieved in a different way compared to a traditional 
cutout service, such as IVOA SIAP\footnote{\href{http://www.ivoa.net/documents/latest/SIA.html}{http://www.ivoa.net/documents/latest/SIA.html}}. 
Using an HTTP-based query syntax, together with 
TCP or UDP based transport protocols, JPIP enables the client to selectively access content of 
interest from the image file, including metadata of interest. This capability results in a vast improvement 
in bandwidth efficiency and speed when performing some very important and valuable image viewing 
tasks in a client/server environment, while reducing the storage and processing requirements of the 
client. The larger the images -- and the more constrained the bandwidth between client and server -- 
the greater are the benefits brought by JPIP.

JPIP clients access imagery on the basis of a so-called ``Window of Interest" (WOI).  The WOI 
consists of a spatial region, at a given resolution, within one or more image components in one 
or more underlying compressed codestreams, optionally limited to a desired quality level or amount 
of communicated data.  In advanced applications, the WOI may also be expressed relative to one or 
more higher level composited representations whose definition depends on metadata.  JPEG2000 
enables the efficient identification and extraction of elements from the compressed codestream(s) 
that intersect with the WOI. This means that from a single  compressed image, a user can remotely 
extract a particular region of the image, a larger or smaller version of the image, a higher or lower 
quality version of the image, or any combination of these. JPIP can be used to progressively forward 
images of increasing quality, giving the client a view of the image as quickly as possible, which 
improves as rapidly as possible, along the direction of interest.

Such features are most desirable for extremely large radio astronomy images, which can hardly 
be used without examining the metadata and previewing the image at low resolution first, transferring 
only the selected parts of the image to a user's computer. This would normally require generating low 
resolution images, thumbnails and metadata and linking them all together in a database. In a system 
equipped with JPEG2000 and JPIP, however, it is only necessary to store a single file per image; 
lower resolutions and thumbnails can be extracted directly out of this high-resolution JPEG2000 
``master" image and streamed or downloaded to the client. This removes the need to store, manage, and link images of 
different resolutions in the database, which can be cumbersome.

In a typical application, when the user chooses to view a particular image, only the resolution layer required 
to view the entire image on the screen need be transferred at first. Quality layers are downloaded progressively 
to give the user an image as quickly as possible. When the user zooms into a particular region of interest in the 
image, only that portion of the image is transferred by the server, and only at the resolution that is of interest. 
Again, the image can be transferred progressively by quality layers. The user can continue to zoom into the image 
until the maximum quality/resolution is reached, and pan across the image; each time, transferred content is limited 
to the area of the image being viewed. An interactive user might then scan across different images of a series, 
maintaining the same region and resolution of interest. Again, only the relevant content is actually transferred. 
The result is a dramatic increase in speed of viewing, and significant increase in the quality and efficiency of the 
viewing experience.

\subsection{JPIP Stream Type}
The JPIP allows three different types of image data to be transmitted between the server and client: 1) full, 
self-contained compressed images (typically, but not necessarily, in the JPEG2000 format); 2) tile data; and 3) 
precinct data \citep{Taubman}.

\textit{Full JPEG2000 Images}. For this data type the server sends the client 
complete JPEG2000 images, at the requested resolution. The resolution level 
is selected to fit in the display window. Because the JPEG2000 images are 
self-contained, they do not require any additional metadata or headers during 
transmission; the images are simply sent to the client and the client decodes them.

\textit{Tiles}. Tiles are rectangular spatial regions within an image that are independently encoded. It can be useful 
to encode a large image as multiple independent tiles, but even huge images can be encoded as a single tile. A tile-based 
JPIP service is useful where numerous small tiles have been used during the encoding process; this allows the server to 
send only the relevant tiles to the client, for decoding. Because tile data is not a self-contained image, additional 
JPIP messaging headers are attached to convey to the client the contents of the messages. Tiling has been used 
in a number of image formats (e.g. TIFF). It has been introduced in FITS along with the compression applied 
to the tiles rather than to the entire image \citep{2010A&A...524A..42P}.
 However, the use of small tiles reduces compression efficiency and can have a large adverse effect upon the 
 service of reduced resolution imagery, since the effective size of the tiles within reduced resolutions can become 
 very small. In JPEG2000 tiles are not considered as a preferred method of structuring an image, as 
 \textit{precincts} offer more advanced solution.

\textit{Precincts}. Precincts are fundamental spatial groupings within a JPEG2000 codestream.  Unlike tiles, 
which represent independently coded regions in space, precincts are defined in the wavelet transform domain. 
The detail subbands at each resolution level are partitioned into independently coded blocks, which are assembled 
into precincts.  Each precinct represents a limited spatial region within the detail bands that are used to augment 
the displayed imagery from a given resolution to the next.  Since precincts are defined in the transform domain, 
their contributions to the reconstructed imagery are overlapping. This means that a server which sends the precincts 
that are relevant to a particular WOI is also sending some content that belongs to surrounding regions, whose 
extent is resolution dependent.  Precincts are the providers of ROI functionality in JPEG2000. The content of a 
precinct can be sent progressively, so as to incrementally refine the quality of the associated imagery.  Additional 
JPIP messaging headers are attached to the precinct data to convey to the client their contents. This image type 
is often the most efficient, as it requires the smallest amount of data to be transmitted; moreover, it is equally 
efficient at all spatial resolutions, unlike tiles, whose size can be optimized only for at a pre-determined resolution. 
An interesting potential mechanism for exploiting precincts within ASKAP and SKA applications, would be to use 
source finding algorithms to automatically generate a catalogue of the most relevant precincts, as part of the 
telescope pipeline. This would enable the selective storage of precinct data based on relevance (from lossy up 
to potentially numerically lossless), as well as the selective delivery of those precincts to a JPIP client; ``empty" 
parts of an image can be sent at much lower quality or resolution, saving the bandwidth, storage/archive space, 
and increasing the speed of fetching and viewing the data. 

\subsection{JPIP Operation and Features}
The client application generates and sends to the server a properly formatted JPIP 
WOI request, containing information about the specific region of the image that the 
user wishes to view, along with the desired resolution, image components of interest 
and optionally explicit quality constraints -- alternatively, the client may request everything 
and expect to receive a response with progressively increasing quality. The JPIP server 
parses the request, calls the JPEG2000 library to extract the relevant image data, and 
sends back to the client a formatted JPIP response. When the response data is received, 
the JPIP client extracts the codestream elements and inserts them into a sparse local 
cache, from which the imagery of interest is decompressed, rendered and/or further 
processed on demand.  Importantly, JPEG2000 codestreams have such a high degree 
of scalability that any image region of interest can be successfully decoded from almost 
any subset of the original content on the server, albeit at a potentially reduced quality.  
This means that decompression and rendering/processing from a local JPIP cache is 
an asynchronous activity that depends only loosely on the arrival of suitable data from 
the server.  To the extent that such data becomes available, the quality of the 
rendered/processed result improves. 

Tile and precinct ``databins" are the basic elements of a JPEG2000 image used by JPIP. 
JPEG2000 files can be disassembled into individual finer elements, called \textit{databins}, 
and then reassembled. Each databin is uniquely identified and has a unique place within 
a JPEG2000 file. Full or partial databins are transmitted from the server to the client in 
response to a JPIP request. The JPIP client can decode these databins and generate 
a partial image for display at any point while still receiving data from the server.

JPIP provides a structure and syntax for caching of databins at the client, and for communication 
of the contents of this cache between the client and the server. A client may wish to transmit 
a summary of the contents of its cache to the server with every request, or allow the server 
to maintain its own model of the client cache by maintaining a stateful session. In either case, 
a well behaved server should reduce the amount of data it is transmitting in response to a JPIP 
request by eliminating the databins that the client has already received in previous transmissions. 
In this way, JPIP provides a very efficient means for browsing large images in a standards-compliant 
fashion.

Both precinct and tile databins have the property that they may be incrementally communicated, 
so that the quality of the associated imagery improves progressively.  JPIP also provides for the 
partitioning of metadata into databins, which can also be communicated incrementally.  This allows 
large metadata repositories to be organised and delivered on demand, rather than as monolithic 
data sets. Moreover, metadata can be used to interpret imagery requests and the image WOI can 
also be used to implicitly identify the metadata that is of interest in response to a JPIP request.

While databins are being transferred between the server and the client, they usually get split 
up into smaller chunks, called \textit{messages}. The JPIP server decides the JPIP message 
size. This flexibility to transmit partial databins enables one to vary the progressive nature of 
the data being sent to the client. If entire databins are sent, first for the lower resolution levels 
in the codestream and then for the higher resolution levels, the imagery pertaining to the 
requested WOI will be received in a resolution-progressive fashion; if messages from different 
databins at the same resolution level are interlaced, the data will be received by the client in 
a quality-progressive order. This flexibility allows applications to control the user experience, 
depending on the application requirements \citep{Taubman}.

There are numerous implementations of JPIP servers and client SDK available:
OpenJPEG JPIP\footnote{\href{https://code.google.com/p/openjpeg/wiki/JPIP}{https://code.google.com/p/openjpeg/wiki/JPIP}},
LEADTOOLS\footnote{\href{http://www.leadtools.com/sdk/jpip/}{http://www.leadtools.com/sdk/jpip/}},
KDU SDK from Kakadu Software\footnote{\href{http://www.kakadusoftware.com/index.php?option=com\_content\&task=view\&id=25}{http://www.kakadusoftware.com}},
2KAN\footnote{\href{http://www.2kan.org/demonstrator.html}{http://www.2kan.org/demonstrator.html}},
JPIPKAK as part of Geospatial Data Abstraction Library\footnote{\href{http://www.gdal.org/frmt\_jpipkak.html}{http://www.gdal.org/frmt\_jpipkak.html}},
and other.

\section{Benchmarking of JPEG2000 compression on radio astronomy images} \label{cha:Bench1}

As yet, JPEG2000 has not been used in astronomy very widely. Most of the accessible radio astronomy images are stored 
in FITS or CASA Image Tables. At the time when this investigation started there was
no software available to convert FITS or CASA Image Tables to JPEG2000 images
with a sufficient range of encoding parameters. To begin with, we limited ourselves to 
encoding FITS images only, as the most common image format currently used in astronomy.

\subsection{Software}
\texttt{f2j} software was developed to convert FITS files to JPEG2000 images.  
The software has been written in C using the open source OpenJPEG\footnote{\href{http://www.openjpeg.org}{http://www.openjpeg.org}} 
codec version 1.0\footnote{v2.0 was already available at the time when the paper was written.} for JPEG2000 compression and NASA's 
\texttt{cfitsio}(\href{http://ascl.net/1010.001}{ascl:1010.001}) 
library for reading FITS files\footnote{\texttt{f2j} does not transfer FITS headers to 
JPEG2000 files, however, the software described in \citealt{2014arXiv1401.7433P} does 
transfer FITS headers into metadata boxes of JPX.}. \texttt{f2j} is an open source software, and can be 
downloaded from the Github\footnote{\href{https://github.com/ICRAR/f2j.git}{https://github.com/ICRAR/f2j.git}}.  

\texttt{f2j} encodes FITS files as JPEG2000 images with a single component consisting 
of greyscale pixel intensities stored as 16 bit unsigned integers.  Each plane 
of a data-cube is written to a separate JPEG2000 image.  \texttt{f2j} reads a full plane from a FITS 
file into an array and then processes each raw value in this array into a greyscale pixel intensity.  
This results in, what is essentially, a bitmap image being passed to the JPEG2000 encoder.  

There are multiple options as to how raw FITS data may be transformed into pixel intensities.  
The particular transformation applied depends on the data type used to store the raw FITS values.  
In the case of 8 or 16 bit integer data, raw values may be used directly as pixel intensities.  
At the time of the first trials OpenJPEG v1.0 codec did not support floating point data directly, so floating point values had to 
be converted to integers in order to create a JPEG2000 image from such data. Later releases of OpenJPEG, 
however, already support a full range of data types that includes double and single precision floating point.

Arbitrary transformations may be defined in \texttt{f2j} for this purpose and it is relatively easy to add 
new transformations to the program.  The floating point transformations currently implemented 
work by assigning the smallest and largest raw data values in the FITS file to the lowest and 
highest possible pixel intensities respectively and then scaling the intermediate data in various ways.  
The logarithmic, square root and power scales are available.  

The JPEG2000 standard specifies that image components may be represented with arbitrary precisions 
up to 38 bits \citep{Schelkens..2009}, however OpenJPEG stores pixel intensities using 32 bit integers 
(in the internal structure it uses to represent an image prior to passing it to the JPEG2000 encoder), 
limiting the precision attainable.  Through experimentation it was also determined that OpenJPEG 
could not correctly encode and decode images using 32 bit precision due to an error in the library. 
However, the used imaged had the dynamic range that could be mostly sufficiently accurately represented 
by 16 bit precision.

In the case of floating point data, it was observed that files would often use only a tiny 
portion of the full range of values supported by this data type.  As the data is scaled to the minimum 
and maximum of the allowable 16 bit integer range, the small range of values being scaled would lessen 
the loss of precision as a result of this quantisation.  

There are many compression options that may be specified affecting how an image is encoded 
using JPEG2000, such as the number of resolutions in the file, tile sizes, compression ratios and 
the use of lossy or lossless compression. \texttt{f2j} supports almost 
all of the compression options supported by OpenJPEG codec v1.0.  

As we were interested in testing the quality of JPEG2000 compression on radio astronomy images,
as well as converting FITS files to JPEG2000, \texttt{f2j} had been equipped with some benchmarking and 
experimentation features. The software is capable of adding varying amounts of Gaussian noise to an image 
to investigate the effects of noise on the compression process. It can perform quality benchmarks 
to examine how lossy compression degrades an image, by decompressing an encoded JPEG2000 
image from a file and calculating quality metrics comparing it to the uncompressed image.

While we have acknowledged the usefulness of JPIP protocol, and it's ability to significantly extend the \emph{cutout}
method of interrogation of image data, which is currently the mainstream method in astronomy, in all
further presented tests in this paper we benchmarked only image files stored on a local drive leaving
benchmarking of JPIP for a future investigation.

\subsubsection {Metrics}

We have built-in \texttt{f2j} the options to calculate the mean squared error (MSE), root mean 
squared error (RMSE), peak signal to noise ratio (PSNR), mean absolute error (MAE), fidelity 
and maximum absolute distortion (MAD) metrics (as well as intermediate data for these metrics). These 
metrics are recommended for compression benchmarks \citep{Delcourt:2011:EFB:2436496.2436501}.  

In practice, we've found the fidelity metric to be unhelpful, as in none of the tests conducted 
did it drop below 0.98, even for badly distorted images.  MSE, RMSE and PSNR are all re-expressions 
of the same information and thus interchangeable. PSNR was found to be the most intuitive to work 
with and was therefore used in most of our tests. 

\begin{figure}
  \centering
  \includegraphics[width=90mm]{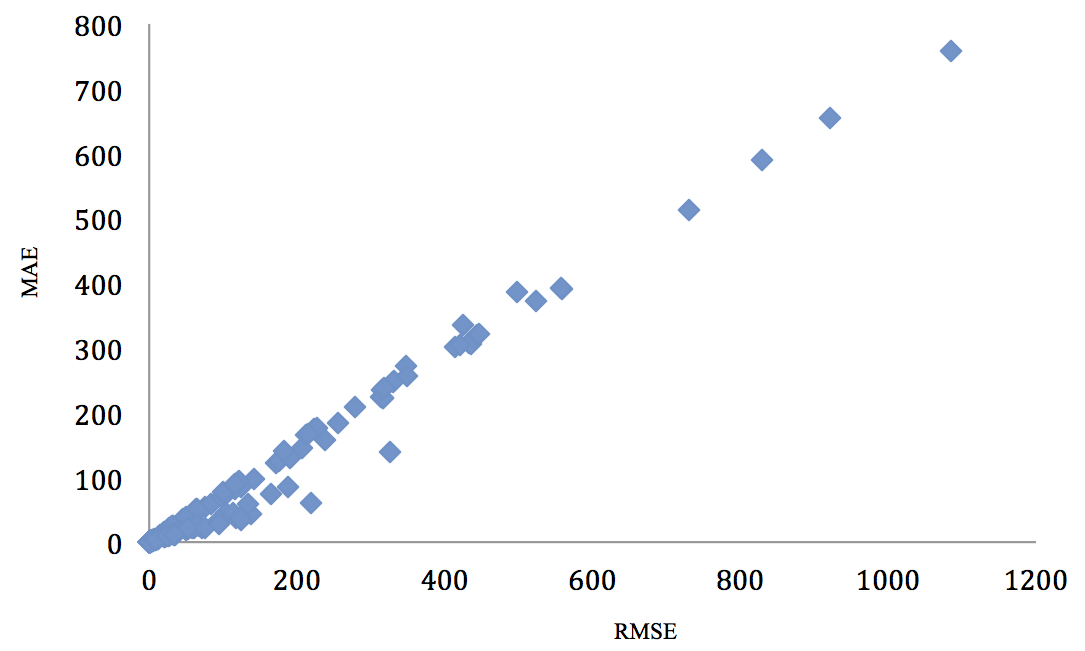}
  \caption{Mean absolute error versus root mean squared error (both axes use units of 16 bit pixel intensities).}
  \label{fig:MAEvsRMSE}
\end{figure}

MAE is not directly related to RMSE, but one would intuitively expect these metrics to be closely correlated.  
This was verified in practice. Figure~\ref{fig:MAEvsRMSE} shows the values of MAE collected in 
quality versus compression ratio benchmark tests (vertical axis) plotted against RMSE values 
(horizontal axis) -- a clear linear relationship is visible. The correlation coefficient between the 
two variables is 0.992.  The close correlation between RMSE and MAE supports the conclusion that 
MAE offers little information beyond RMSE (and thus PSNR).  Thus our discussions of results will 
focus on PSNR and MAD mostly. 

\subsubsection{Test images}
A large number of publicly available radio astronomy FITS files were examined to come up with a representative set of 
test images representing features and attributes that radio astronomers would expect to encounter.  
These include sparsely and densely populated images, dominant and diffuse features, high or low 
noise and regular or random noise.  A final test set of 11 images was selected, including 9 planar 
images and 2 data cubes.  All images contained floating point data. These images were used in the 
benchmarking described in the following sections.

\subsection{Lossless compression benchmarking}
These benchmarks involved encoding the test images losslessly and observing 
the compression ratios attainable.  

There are many parameters that might be specified when encoding to an JPEG2000 image.  
These allow the image compression to be fine-tuned for a particular purpose, \emph{i.e.} for 
distribution as part of a JPIP system and have the potential to affect compression ratios and 
image quality (for lossily encoded images).  For the initial benchmarking, other than altering the 
compression ratios and target quality for the lossy compression benchmarks (below), the 
default OpenJPEG settings were used as typical parameters. Therefore, while our project 
provides a guide to the compression performance possible using JPEG2000, best results for 
any practical application will result from optimising the compression process for a particular use.  

\begin{table*}[!t]
\centering
\resizebox{18cm}{!} {
\begin{tabular} { l || r | r | c | c}
\hline
File & \parbox[t]{2.7cm}{Size of JPEG2000 file (bytes)} &  \parbox[t]{2.7cm}{Size of FITS file (bytes)} & \parbox[t]{2cm}{Compression Ratio} & Disk Space Saved (\%)\\
\hline
1.45I1.50\_AM0381\_1992DEC14\_1\_125.U50.7S.imfits  & 110,689 & 406,080 & 1:3.67 & 73\\
1.45I4.68\_AK456\_1998AUG28\_1\_76.1U2.95M.imfits & 120,021 & 406,080 & 1:3.38 & 70\\
1.45I4.70\_AK456\_1998SEP04\_1\_131.U1.68M.imfits	& 86,841 & 406,080 & 1:4.68 & 79 \\
1.45I6.65\_TESTS\_1994JUL24\_1\_120.U2.61M.imfits	 & 74,199 &406,080 & 1:5.47 & 82\\
1.45I9.04\_AB778\_1996JAN29\_1\_42.6U4.91M.imfits	 & 103,796 &406,080 & 1:3.91 & 74\\
1.45I10.1\_AK456\_1998NOV15\_1\_23.3U4.63M.imfits & 118,756 & 406,080 & 1:3.41 & 71\\
00015+00390Z.fits & 2,459,591 & 7,145,280 & 1:2.90 & 66\\
22.4I0.94\_AF350\_1998DEC24\_1\_3.41M55.7S.imfits	 & 209,432 & 898,560 & 1:4.29 & 77\\
CYG.ICLN.FITS & 1,696,799 & 16977600 & 1:10.01 & 90\\
M31\_5Mpc\_dirty\_6km.fits & 635,258,960 & 2,073,605,760 & 1:3.26 & 69\\
M31\_model\_5Mpc.fits & 12,854,386	& 652,916,160 & 1:50.79 & 98\\
\hline
\end{tabular}}
\caption{Lossless compression benchmarking results. The compression ratios are true ratios 
for all the images including those that had been truncated from 32 bit to 16 bit integers.}
\label{tab:LL_br}
\end{table*}

\begin{table*}[!t]
\centering
\resizebox{15cm}{!} {
\begin{tabular} { l || l | c | c}
\hline
\parbox[t]{2cm}{Compression Ratio} & File & \parbox[t]{2.7cm}{PSNR (bB)} &  \parbox[t]{2.7cm}{MAD}\\
\hline
1:15 & 1.45I1.50\_AM0381\_1992DEC14\_1\_125.U50.7S.imfits	& 46.3 	& 2241\\
	& 1.45I4.68\_AK456\_1998AUG28\_1\_76.1U2.95M.imfits	 	& 41.4	& 3295\\
	& 1.45I10.1\_AK456\_1998NOV15\_1\_23.3U4.63M.imfits	 	& 42.0	& 3590\\
	& 00015+00390Z.fits									& 45.5	& 2031\\
	& M31\_5Mpc\_dirty\_6km.fits (110)						& 49.2	& 3942\\
\hline
1:20	& 1.45I9.04\_AB778\_1996JAN29\_1\_42.6U4.91M.imfits		& 46.3	& 1859\\
	& M31\_5Mpc\_dirty\_6km.fits (40)						& 47.4	& 2621\\
	& M31\_5Mpc\_dirty\_6km.fits (75)						& 46.4	& 3941\\
\hline
1:25	& 1.45I4.70\_AK456\_1998SEP04\_1\_131.U1.68M.imfits		& 51.4	& 1760\\
	& M31\_5Mpc\_dirty\_6km.fits (5)						& 43.4	& 3792\\
	& M31\_5Mpc\_dirty\_6km.fits (145)						& 43.8	& 3932\\
\hline
1:30	& 1.45I6.65\_TESTS\_1994JUL24\_1\_120.U2.61M.imfits		& 54.5	& 1987\\
	& 22.4I0.94\_AF350\_1998DEC24\_1\_3.41M55.7S.imfits		& 48.8	& 3160\\
\hline
\end{tabular}}
\caption{Quality benchmarks for lossy compression at the compression ratio of first visual degradation. The compression ratios in the 
table are those supplied to the JPEG2000 encoder. The numbers in brackets next to the file names 
indicate a particular plane (frequency channel) of a data cube. }
\label{tab:QB}
\end{table*}

Table~\ref{tab:LL_br} shows the lossless compression ratio attained for each of the 11 test 
images, the space saved as a result of compression, and the sizes (in bytes) of the JPEG2000 images 
and original FITS files.  

\begin{figure}
  \centering
  \includegraphics[width=90mm]{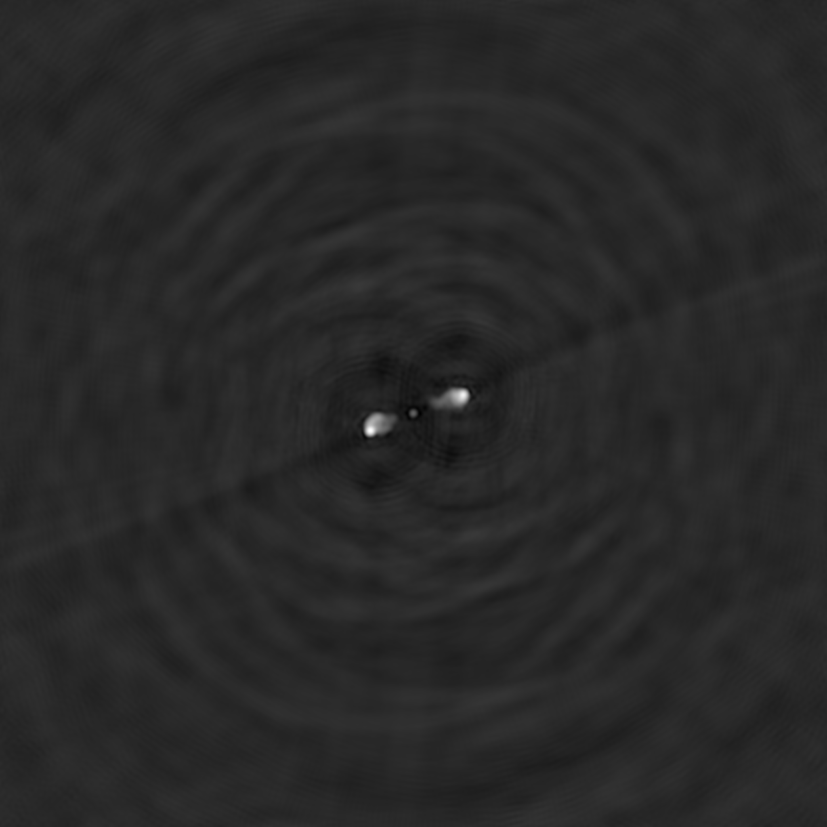}
  \caption{Cygnus A as observed on the EVLA converted to JPEG2000.}
  \label{fig:CygA}
\end{figure}

In terms of lossless compression ratios, there are two obvious outliers in this table: 
M31\_model\_5Mpc.fits and CYG.ICLN.FITS.  The first is a very clean data cube 
containing a simulated ASKAP image of the M31 
galaxy\footnote{\href{http://www.atnf.csiro.au/people/Matthew.Whiting/ASKAPsimulations.php}{http://www.atnf.csiro.au/people/Matthew.Whiting/ASKAPsimulations.php}}, 
which achieved a compression ratio of 1:50.79.  The second image is of Cygnus A 
observed on the EVLA\footnote{Credit to Richard Dodson of ICRAR for the original FITS file}, 
which achieved a more modest compression ratio of 1:10.01 
(see Figure~\ref{fig:CygA}). This image contained a reasonable amount of instrumental 
noise, but nevertheless this noise could be represented efficiently using the JPEG2000 
lossless algorithm.

\begin{figure}
  \centering
  \includegraphics[width=90mm]{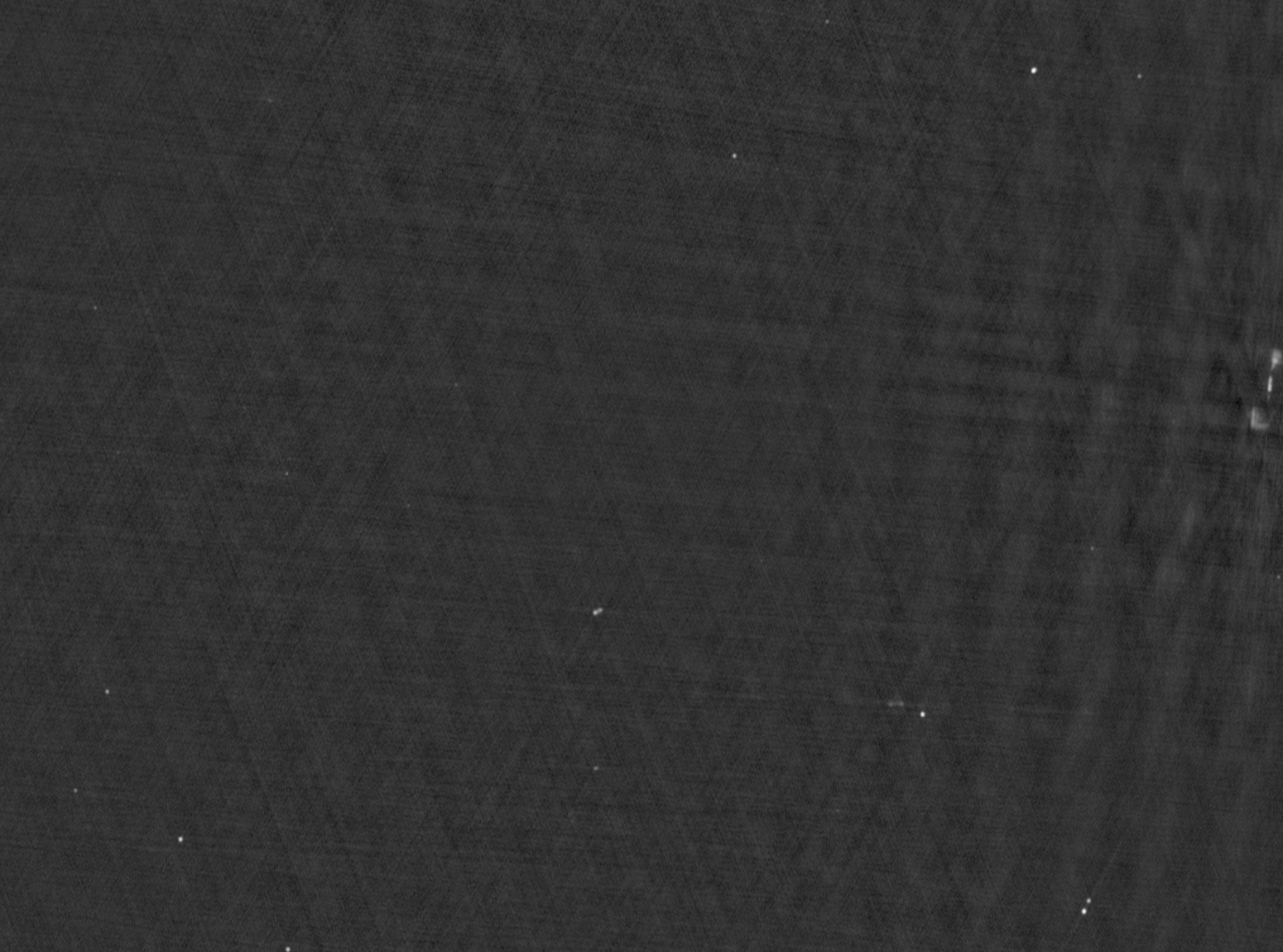}
  \caption{00015+00 as observed on the VLA array converted to JPEG2000.}
  \label{fig:VLA00015}
\end{figure}

Not taking in account the outliers, the mean compression ratio was 1:3.89 with a standard deviation of 0.80.  
Of the remaining images, the worst compression ratio, of 1:2.90, occurred with the file 00015+00390Z.fits 
(see Figure~\ref{fig:VLA00015}). This was a very noisy (mostly instrumental) image as observed on the VLA array. 
The worst compression ratio this image was achieved despite the fact that the instrumental noise has a regular 
but very finely gridded structure.

\begin{figure}
  \centering
  \includegraphics[width=90mm]{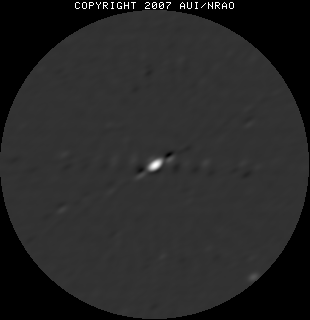}
  \caption{RC2357 as observed on the VLA array.}
  \label{fig:RC2357}
\end{figure}

Of the remaining files, the best compression ratio of 1:5.47 was achieved on the file 
1.45I6.65\_TESTS\_1994JUL24\_1\_120.U2.61M.imfits, which contained a relatively 
clean image of RC2357 as observed from the VLA array (see Figure~\ref{fig:RC2357}).
The image also has constant values in all four corners that contributes to the high
compression ratio.

In all of our test cases lossless compression gave a significant disc space saving (see Table~\ref{tab:LL_br}).

\subsection {Lossy compression benchmarking}
\subsubsection {Quality versus compression ratio benchmarks}
These benchmarks involved compressing the test images lossily by specifying a particular 
compression ratio to the JPEG2000 encoder.  Compression and quality metrics were recorded 
for each of the compressed images. The compression ratios at which compression artefacts first 
became visually noticeable (relative to the losslessly compressed version) were recorded. 
The residual images resulting from the lossy compression process were written to files 
and were visually examined for features of interest. 

Compression ratios of 1:$X$ were used, where $X$ took the values \emph{25, 20, 15, 10, 5, 2, 1.5}.  
Higher compression ratios were examined if there were no visible compression artefacts at the 
1:25 compression ratio.  

Table~\ref{tab:QB} shows the nominal compression ratio at which each file first showed 
\emph{visual} degradation and quality metrics at this point. Note that the compression ratios in the 
table are those supplied to the JPEG2000 encoder. The numbers in brackets next to the file names 
indicate a particular plane (frequency channel) of a data cube. 

The first point to note is that every file could be compressed lossily to a nominal 1:10 ratio without showing 
visual degradation, which is 2.6 times greater than the average 1:3.89 compression ratio attainable 
using lossless compression.  

\begin{figure}
  \centering
  \includegraphics[width=90mm]{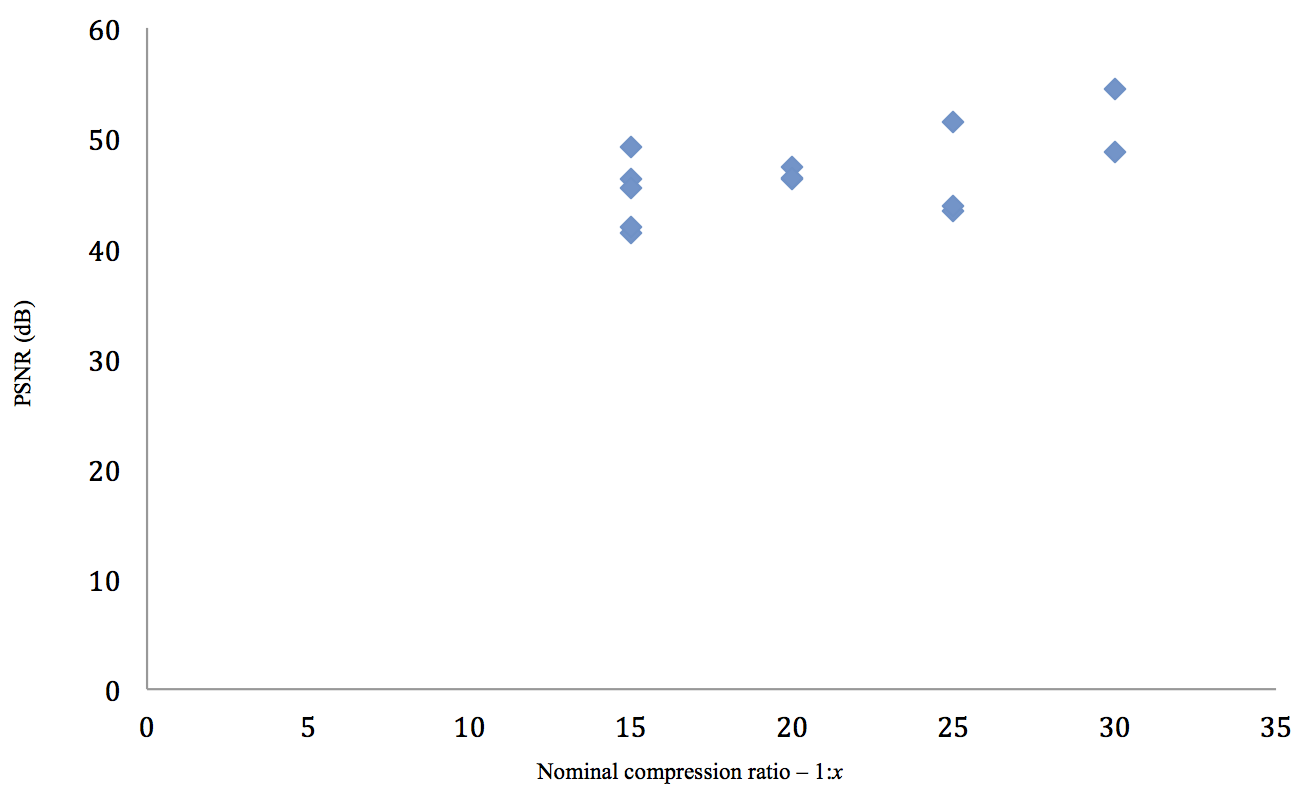}
  \caption{PSNR at the nominal compression ratio at which visual degradation first occurred.}
  \label{fig:PSNRdegrad_2}
\end{figure}

Figure~\ref{fig:PSNRdegrad_2} shows the PSNR values recorded at the compression ratios that visual 
degradation first occurred.

From the graph, it is obvious that while visual degradation first occurred over a 
relatively narrow PSNR range, it occurred over a relatively wide nominal 
compression ratio range. This observation motivated the next set of benchmarks.  

\subsubsection{Compression ratio versus quality benchmarks}

These benchmarks investigated the opposite side of the equation to the previous set of benchmarks.  
The tests proceeded as in the previous section, except that the test images were compressed lossily 
by specifying a particular target quality, expressed as a peak signal to noise ratio (PSNR), to the 
JPEG2000 encoder, rather than specifying a compression ratio. For these benchmarks, FITS files 
were encoded lossily with a particular targeted quality (PSNR).  
The quality metrics here are thus largely influenced by this compression parameter -- therefore 
it is the compression metrics that are of interest in these tests.  

\begin{table*}[!t]
\centering
\resizebox{12cm}{!} {
\begin{tabular} { l | l | l }
\hline
\parbox[t]{2.7cm}{PSNR (bB)}  & File & \parbox[t]{2cm}{Compression Ratio}\\
\hline
50	&	1.45I1.50\_AM0381\_1992DEC14\_1\_125.U50.7S.imfits	&	1:24\\
	&	1.45I4.70\_AK456\_1998SEP04\_1\_131.U1.68M.imfits		&	1:62\\
	&	1.45I6.65\_TESTS\_1994JUL24\_1\_120.U2.61M.imfits		&	1:80\\
	&	1.45I9.04\_AB778\_1996JAN29\_1\_42.6U4.91M.imfits		&	1:32\\
	&	22.4I0.94\_AF350\_1998DEC24\_1\_3.41M55.7S.imfits		&	1:58\\
40	&	1.45I4.68\_AK456\_1998AUG28\_1\_76.1U2.95M.imfits	&	1:41\\
	&	1.45I10.1\_AK456\_1998NOV15\_1\_23.3U4.63M.imfits	&	1:36\\
	&	00015+00390Z.fits	&	1:111\\
	&	M31\_5Mpc\_dirty\_6km.fits (5)\\
	&	M31\_5Mpc\_dirty\_6km.fits (40)\\
	&	M31\_5Mpc\_dirty\_6km.fits (75)\\
	&	M31\_5Mpc\_dirty\_6km.fits (110)\\
	&	M31\_5Mpc\_dirty\_6km.fits (145)		&	1:102\\
\hline
\end{tabular}}
\caption{Compression benchmarks at the quality (PSNR) of first visual degradation.}
\label{tab:QB_PSNR}
\end{table*}

Table~\ref{tab:QB_PSNR} shows the compression ratios achieved at the PSNR (quality) 
that files first showed visual degradation.  

\begin{figure}
  \centering
  \includegraphics[width=90mm]{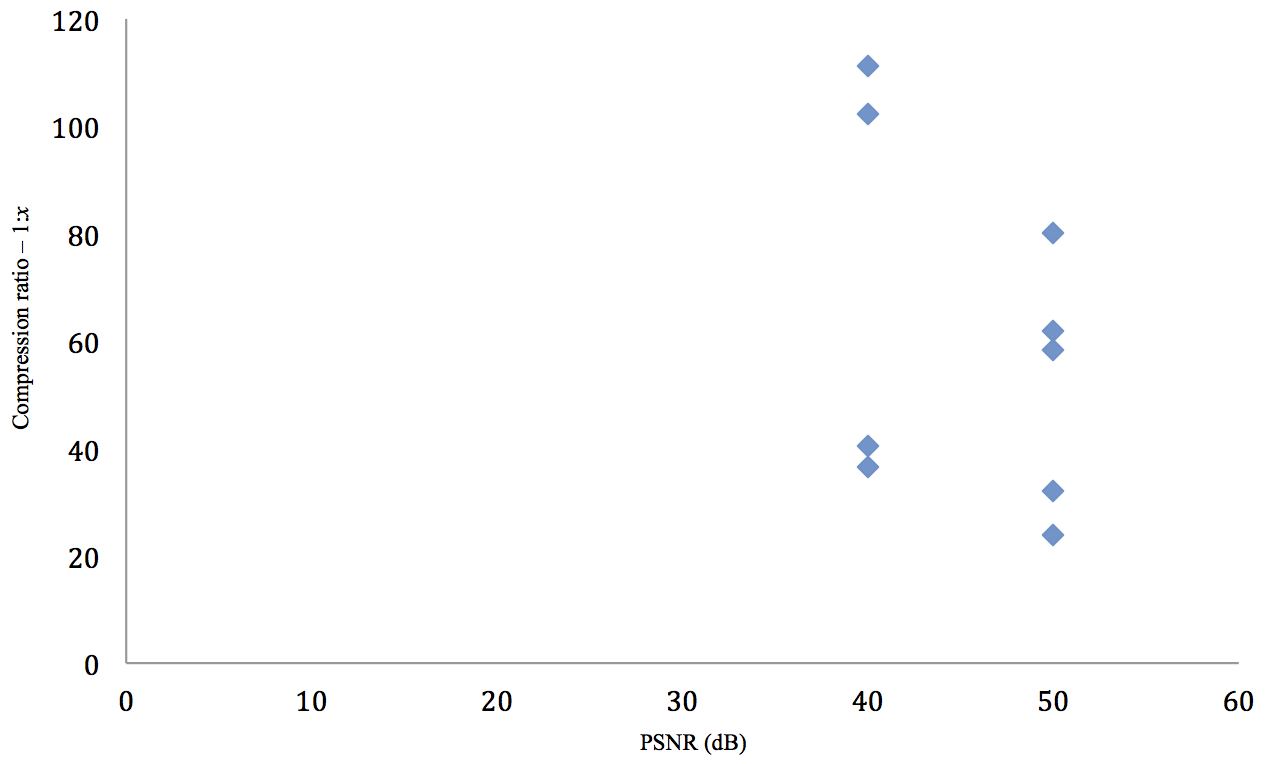}
  \caption{Compression ratio at the PSNR at which visual degradation first occurred.}
  \label{fig:degrad_vis_M51less}
\end{figure}

Figure~\ref{fig:degrad_vis_M51less} shows the compression ratios attained 
at the PSNR that visual degradation first occurred.  Again, it is obvious that visual degradation first occurs 
over a relatively wide range of compression ratios but a relatively narrow PSNR range.  Thus the 
target PSNR appears to predict visual image quality far better than compression ratio.  When working 
with lossily compressed images, it would thus be advisable to encode images for a particular quality 
using PSNR as a metric rather than a particular compression ratio.

\section{Improving existing formats, adopting other technologies or starting from scratch?} \label{cha:adopting}
\subsection{Improving of existing formats}
The FITS standard, in its present form, is clearly unable to support such cases as ASKAP or SKA.
Can it be improved? \citealt{white2012tiled} have made perhaps the most significant attempt 
to improve FITS for the large images use case. The convention suggests that compressed tiles of image 
are stored in a binary table extension which is hidden from the end-user as the image is 
accessed through the same image interface that is used to access normal raw images. However,
this convention does not offer any new framework to work with the imagery data. As before,
a cutout needs to be produced as a separate file and downloaded to to the client. Comparing to the
described JPIP client-server framework the cutout framework is clearly limiting for many use cases, especially 
that involve visualisation.

Another problem for improving existing standards such as FITS, is its expected and important property 
of backwards compatibility. Such legacy often conflicts with the modern 
performance and flexibility requirements \citep{2011ASPC..442...53A}. On the other hand, FITS rather loosely specifies
the formats for various uses, and conservatively defines the metadata. As the result, the actual use of FITS 
often deviates from the original specifications in order to accommodate the specific needs of 
projects or to extend the functionality in general. This creates an illusion of a standard, while in reality
there are many proprietary cases of FITS that software can not universally interpret.

These are factors, that, in our view, significantly limiting the opportunity to improve FITS in particularly
to address large data issue. This looming predicament is especially relevant to such projects as SKA, 
wherein certain operational modes will be generating datasets comprising tens of terabytes of individual
data products.

\subsection{Adopting technologies vs starting from scratch}
A standard like JPEG2000 requires many years of development by the top experts in the field. 
The amount of investment in both, time and money required to implement the standard are much greyer.
Many widely adopted standards are often supported by both commercial and open source developments. 
The downsides are that not all standards are royalty free to use in development, and that 
it might be more difficult to influence the development 
of a standard to accommodate the needs of rather small astronomy community. 
The industry interest to collaboration in developing standards can be piqued by the high public profile 
astronomy projects that can be used as a vehicle for promotion of a standard 
or technology. Projects like SKA may represent the unique opportunities for effective collaboration 
between the astronomy community and the industry R\&D.

Clearly, no single industry standard/technology can address all the needs of astronomy. JPEG2000 
is not universal, and only limitedly suitable for handling other types of data e.g. visibilities 
for radio interferometers. The functionality required for other types of data is significantly different to the 
functionality required for visual exploration of image data. However, we considered JPEG2000 
in detail to demonstrate how powerfully an industry standard can address the requirements for 
large astronomical imagery. Use of a suitable standard opens access to many tools that are 
readily available providing a shortcut to the solutions that would take many years to achieve 
otherwise.

Moreover, we would like to argue that the ability to exchange and correctly interpret the
data is more important than: the format of data that needs to be optimised; a particular 
use case; or an optimisation to a hardware platform. As long as there is a clear description of data
in an universal way, the data can be extracted from or ported to any particular format as necessary.

We would like to urge the astronomical community to begin the work of defining the new set of standards
to provide a guidance for new developing instruments to efficiently store and exchange the staggering 
amount of data that are going to be generated in the next decade. The work that has been done by IVOA
over the last decade can be a great asset.

\section{Conclusion} \label{cha:conc}
New telescopes, such as the SKA, will produce images of extreme sizes. Providing adequate 
performance and level of convenience when serving such images to the end-user is going 
to be beyond the capabilities of current astronomy image formats. Improvements 
of the existing image and data formats can not solve the deficiencies inherently there, due to the fundamental 
limitations at the time of development. 

New advanced technologies
are necessary. Technologies such as JPEG2000 have the potential to powerfully pave the way to 
a contemporary solution that will adequately address the challenges of extremely large imagery. 

Substantial reductions in storage/archive requirements can be achieved by losslessly encoding data into 
JPEG2000 images. Even greater saving may be achieved through lossy compression.  

JPIP provides a standard powerful way for interaction with the imagery data reducing the bandwidth, 
storage, and memory requirements, and increasing the mobility of future astronomy application.

The results of our benchmarks demonstrate the viability of JPEG2000 compression for storing 
and distributing radio astronomy images.  JPEG2000 is not just about compression -- it has 
the potential to enable an entirely new paradigm for working with radio astronomy imagery data.  

\bibliography{references}

\end{document}